\documentclass[]{JHEP}
\usepackage{epsfig}
\usepackage{cite}

\title{Detecting Exotic Heavy Leptons at the Large~Hadron~Collider}

\author{B.C.~Allanach$^{\ast}$, C.M.~Harris$^{\dagger}$, 
M.A.~Parker$^{\dagger}$, P.~Richardson$^{\dagger,\ddagger}$ and 
B.R.~Webber$^{\dagger}$\\
$^{\ast}$Theory Division, CERN, 1211 Geneva 23, Switzerland. \\
$^{\dagger}$Cavendish Laboratory, University of Cambridge, Madingley Road, 
Cambridge, CB3~0HE, UK.\\
$^{\ddagger}$DAMTP, Centre for Mathematical Studies, Wilberforce Road, 
Cambridge, CB3~0WA, UK.\\
}

\abstract{New almost-degenerate charged and neutral heavy leptons are a 
feature of a number of theories of physics beyond the Standard Model.  The 
prospects for detecting these at the Large Hadron Collider using a 
time-of-flight technique are considered, along with any cosmological or 
experimental constraints on their masses.  Based on a discovery criterion of 
10 detected exotic leptons we conclude that, with an integrated luminosity of 
100~fb$^{-1}$, it should be possible to detect such leptons provided their 
masses are less than 950~GeV\@.  It should also be possible to use the angular
distribution of the produced particles to distinguish these exotic leptons 
from supersymmetric scalar leptons, at a better than 90\% confidence level, 
for masses up to 580~GeV\@.
}


\preprint{Cavendish-HEP-01/10 \\
		DAMTP-2001-71 \\
		CERN-TH/2001-205}

\begin{document}

\section{Introduction}

One of the most widely considered possibilities for new physics beyond the 
Standard Model (SM) is supersymmetry (SUSY) \cite{bailin,martin}\@.  
Supersymmetry is theoretically attractive and is able to solve some of the 
problems faced when trying to understand how fundamental models at higher 
energy scales are consistent with what is seen by experiments at low energy 
scales.  SUSY solves the so-called hierarchy problem and improves the 
unification of the three gauge couplings at a higher energy scale.

The simplest possible SUSY extension of the Standard Model is the Minimal 
Supersymmetric Standard Model (MSSM)\@.  The model has a second Higgs scalar
doublet and for each SM particle there is an associated superpartner.

We know that supersymmetry is broken because we do not see 
associated superpartners with the same masses as the known particles.  It has 
long been known that a viable way to break supersymmetry is in a hidden sector
- the SUSY breaking in a particular model is then transmitted to the ``outside
world'' by some means (usually gravitational or gauge interactions).

One scenario motivated by string theory is that of an intermediate scale 
\cite{karim,strings,type1,dbrane}.  This is a result of the realisation that, 
if a fundamental theory includes higher dimensional objects like D-branes, the
fundamental scale is effectively a free parameter and does not have to be 
close to the Planck scale as previously thought.  The fundamental scale could,
in theory, be as low as experimental limits allow (\emph{i.e.} 1~TeV)\@.

In the type of model being considered here, gauge coupling unification is 
assumed to occur at an intermediate scale.  ``Intermediate'' refers to an 
energy scale ($\sim 10^{11}$~GeV) which is the geometric mean of the weak and 
Planck energy scales.  This choice is motivated by the scale of supersymmetry 
breaking in hidden sector, gravity mediated, scenarios.  One of
the strongest arguments supporting the intermediate string scale scenario is 
that of the axion decay constant \cite{strings}.  The axion field is the 
Goldstone mode of a chiral symmetry introduced to the Standard Model to solve 
the strong Charge Parity problem (which gives an unnaturally small bound on 
one of the parameters) \cite{PQ}.  Astrophysical and cosmological bounds 
constrain the axion decay constant to be in the approximate energy range 
$10^9$--$10^{12}$~GeV.  The intermediate scale is also consistent with some 
neutrino mass mechanisms, and is favoured by certain cosmological inflation 
models.  High energy cosmic ray observations and certain non-thermal dark 
matter candidates are also supporting arguments for the intermediate 
energy scale.

Various intermediate scale models are motivated from different string models
in references \cite{karim,strings,type1,dbrane}.  Some phenomenological 
constraints on intermediate scale models are provided by \cite{fernando}, and 
$g-2$ constraints have also been discussed \cite{g21,g22}.

One of the ways of achieving this intermediate scale unification is to include
new leptons as part of extra supermultiplets added to the MSSM\@.  Although 
this choice is not unique, the extra leptons in \cite{fernando} are three 
vector-like copies of left-handed SU(2) doublets ($L^\prime$, 
$\bar{L}^\prime$) and two right-handed singlets ($E^\prime$,$\bar{E}^\prime$).

By assumption there are no new Yukawa couplings for the model in 
\cite{fernando}.  This means that the lightest heavy
lepton is stable and the others will decay into it.  This is possible because
the tree-level mass degeneracy of the charged and neutral leptons will be 
destroyed by electroweak symmetry breaking.
  
As in the MSSM, the renormalisation group equations can be used to determine 
the spectrum of supersymmetric particles which will have phenomenological 
implications for the Large Hadron Collider (LHC).  However the characteristic
feature of this model is the new leptons and so this work concentrates on 
the phenomenology due to the extra leptons themselves.  More specifically we
consider whether it will be possible to detect them at the LHC, and for what 
range of masses.    

Previous limits on the masses of any new quasi-stable charged leptons have 
been determined by the energy available for particle production in lepton 
colliders.  The gauge unification arguments in intermediate scale models mean 
that the extra leptons are expected to have masses in the TeV range.  At the 
LHC enough energy should be available to produce and detect these exotic heavy
charged leptons.

In this paper we first review the existing cosmological and experimental 
limits on heavy particles to ensure that the leptons incorporated in this 
intermediate scale model are not ruled out.  In section \ref{HERWIG} we 
describe the changes to the {\small HERWIG} Monte Carlo event generator to 
take account of these new heavy leptons.  Then we
consider the detection of these leptons at the LHC using a ``time-of-flight'' 
technique.  Heavy leptons will arrive at the detector significantly later than
relativistic particles - in section \ref{impl} the prospects of using this 
time delay as a method of detection are studied for a range of masses.  We 
also consider how to distinguish such leptons from scalar leptons on the basis
of their different angular distributions.  Our results are presented in 
section~\ref{results}.

\section{Cosmological and Experimental Constraints}
\label{litrev}

The mass degeneracy of the charged and neutral leptons will be broken by 
radiative corrections.  The electromagnetic self-energy correction for the 
charged lepton ensures that its mass will be greater than that of the neutral 
lepton\cite{Sher1}.  This means that the charged lepton can decay to its 
neutral partner producing either a real or virtual W, depending on the mass 
difference.  The neutral heavy lepton can be assumed to be totally stable 
(\emph{i.e.} it has a lifetime orders of magnitude longer than that of the 
universe) because of the assumption that there are no new Yukawa couplings.

There are a number of cosmological and experimental limits on leptons 
(particularly charged leptons) as a function of both mass and lifetime:

\begin{itemize}

\item
The relic abundance of the leptons (calculated from the self-annihilation 
cross section) must not ``over-close'' the universe \emph{i.e.} provide more 
than the critical energy density ($\sim 10^{-5}$ GeV cm$^{-3}$) which is 
presumed to be accounted for by Dark Matter\cite{wolf,griest}.

\item
A stable, charged lepton must have a low enough relic abundance for it not to 
have been detected in searches for exotic heavy isotopes in ordinary matter 
(see \emph{e.g.} \cite{water}).

\item
The massive lepton and any decays it may have must not significantly affect
nucleosynthesis or the synthesised elemental abundances \cite{ellis,subir}.

\item
If the lepton decays before the recombination era (at $\sim 10^{12}$~s) it 
must not distort the cosmic microwave background radiation \cite{ellis}.

\item
If the lepton has a longer lifetime it must not contradict the limits from
gamma-ray and neutrino background observations \cite{kribs,gond}.

\item
The mass and lifetime of the new leptons must not be such that they would 
have been detected in a previous collider experiment \cite{LEP}.

\end{itemize}

Most of these limits are summarised in \cite{Sher2} and no constraints on 
charged leptons are found for lifetimes less than $\sim 10^{6}$~s, even for 
masses up to the TeV scale.  The mass splitting between the neutral and 
charged leptons will be of order MeV \cite{Sher1} which means that the 
lifetime of the charged lepton will certainly be shorter than 10$^{6}$~s.

This means the only relevant limits on the mass of the heavy leptons being 
studied are the experimental ones, which rule out masses of less than 
93.5~GeV \cite{LEP}.

\section{Theoretical Input to {\small HERWIG}} 
\label{HERWIG}

The general purpose Monte Carlo event generator {\small HERWIG 6.3} includes 
subroutines for both neutral and charged Drell-Yan processes in a hadron 
collider \cite{HERWIG,HERWIG2,HERWIG3}.  As described in \cite{HERWIG2} the 
initial-state parton showers in Drell-Yan processes are matched to the exact 
$\mathcal{O}(\alpha_S)$ matrix-element result \cite{mike}.  However the 
neutral Drell-Yan processes all use the approximation that the two fermions 
produced can be treated as massless.  It is reasonable that this is a valid 
approximation for the SM quarks and leptons but this will not necessarily be 
the case for the proposed new heavy leptons discussed above which may have 
masses of order 1~TeV\@.

For this reason full Born-level expressions for the Drell-Yan cross sections 
were derived, taking into account the masses of the produced leptons.  This 
was done for the neutral and charged current cases (in both cases the result 
was expressed in terms of vector and axial couplings to keep it as general as 
possible and allow for new couplings in the intermediate scale model).

The derived expressions for both differential and total cross sections for 
neutral current Drell-Yan processes 
($q\bar{q}\rightarrow Z^0/\gamma\rightarrow L^-L^+$) are shown below.  The 
notation is influenced by that already used in subroutines within 
{\small HERWIG 6.3}.

The differential cross section is given by

\begin{equation}
\frac {d\hat{\sigma}}{d\Omega} (q\bar{q}\rightarrow L^-L^+) 
= \frac{e^4}{48\pi^2} \frac {1} {\hat{s}^2} \frac {p_3} {E}
\left[ C_1 \left( E_3E_4 + p_3^2\cos^2\theta^*\right) + C_2 m_3 m_4 +
2 C_3 E p_3 \cos\theta^* \right],
\label{dxs}
\end{equation}

\vspace{2mm}
\noindent where

\begin{equation}
C_1 = \frac {\left[\left(d_V^f\right)^2 + \left(d_A^f\right)^2 
\right] \bigg[ \left(d_V^i\right)^2 + \left(d_A^i\right)^2 \bigg]\hat{s}^2 } 
{\left(\hat{s}-m_Z^2\right)^2 + m_Z^2\Gamma_Z^2} + \left(q^f\right)^2
\left(q^i\right)^2 + \frac {2 q^f q^i d_V^f d_V^i\hat{s}
\left(\hat{s}-m_Z^2\right)}{\left(\hat{s}-m_Z^2\right)^2 + m_Z^2\Gamma_Z^2}, 
\end{equation}

\begin{equation}
C_2 = \frac {\left[\left(d_V^f\right)^2 - \left(d_A^f\right)^2 
\right] \bigg[ \left(d_V^i\right)^2 + \left(d_A^i\right)^2 \bigg]\hat{s}^2 } 
{\left(\hat{s}-m_Z^2\right)^2 + m_Z^2\Gamma_Z^2} + \left(q^f\right)^2
\left(q^i\right)^2 + \frac {2 q^f q^i d_V^f d_V^i\hat{s} 
\left(\hat{s}-m_Z^2\right)}{\left(\hat{s}-m_Z^2\right)^2 + m_Z^2\Gamma_Z^2},
\end{equation}

\begin{equation}
C_3 = 2 \left( \frac {2 d_V^f d_A^f d_V^i d_A^i\hat{s}^2} 
{\left(\hat{s}-m_Z^2 \right)^2 + m_Z^2\Gamma_Z^2} + 
\frac {q^f q^i d_A^f d_A^i \hat{s} \left(\hat{s}-m_Z^2\right)} 
{\left(\hat{s}-m_Z^2\right)^2 + m_Z^2\Gamma_Z^2} \right).
\end{equation}

\vspace{2mm}

\noindent In these expressions $q^i$ and $q^f$ are the charges (in units of
the electron charge) of the initial- and final-state particles respectively.  
The angle $\theta^*$ is the angle between the outgoing lepton ($L^-$) 
direction and the incoming quark direction in the centre-of-mass frame.  
$E_3$ and $p_3$ refer to the centre-of-mass energy and magnitude of momentum
of the produced lepton.  Similarly the subscript 4 refers to the produced
anti-lepton. $E$ is the energy of both the colliding quarks in the 
centre-of-mass frame (\emph{i.e.} $\hat{s}=4E^2$).  

The couplings $d_V$ and $d_A$ are related to the normal vector and axial 
coupling constants to the $\mbox{Z}^0$ ($c_V$ and $c_A$) by relations like

\begin{equation}
d_V = \frac {c_V g_Z} {2e}.
\end{equation}

\noindent Equation~(\ref{dxs}) is a general expression which includes the 
$\mbox{Z}^0$/$\gamma$ interference terms.  The massless case is retrieved by
setting $p_3=E_3=E_4=E$ (as well as $m_3=m_4=0$).  The normal charged
current Drell-Yan case ($q\bar{q}^\prime \rightarrow W^\pm\rightarrow 
L^\pm L^0$) is given by setting $c_V=c_A=1$ and replacing $g_Z$,
$m_Z$, $\Gamma_Z$ and $q^{i/f}$ by $\frac{g_W} {\sqrt{2}}$, $m_W$, 
$\Gamma_W$ and 0 respectively.

Integration of eq.~(\ref{dxs}) gives the following equation for the cross 
section:

\begin{equation}
\hat{\sigma} (q\bar{q}\rightarrow L^-L^+)= 
\frac {e^4} {12\pi} \frac {1} {\hat{s}^2} \frac {p_3} {E} 
\left[ C_1 \left(E_3E_4 + \frac {p_3^2} {3}\right) + C_2 m_3 m_4 \right].
\label{xs}
\end{equation}

\vspace{2mm}

In order to distinguish these heavy leptons from heavy supersymmetric 
partners of SM particles it would be necessary to consider the angular 
distribution of the produced particles.  For the neutral current production of
left/right-handed scalars the expression for the differential cross section 
would be

\begin{equation}
\frac {d\hat{\sigma}}{d\Omega}(q\bar{q}\rightarrow\tilde{l}_{L/R}
\tilde{l}^*_{L/R})=\frac {e^4} {96\pi^2} D \frac {1}{\hat{s}^2} \frac {p_3} 
{E} p_3^2 \sin^2\theta^*,
\label{sdxs}
\end{equation}

\vspace{2mm}
\noindent where

\begin{equation}
D = \frac {h_{L/R}^2 \bigg[\left(d_V^i\right)^2 +\left(d_A^i\right)^2 \bigg] 
\hat{s}^2} {\left(\hat{s}-m_Z^2\right)^2 + m_Z^2\Gamma_Z^2} + 
\left(q^f\right)^2 \left(q^i\right)^2 +
\frac {2 q^f q^i h_{L/R} d_V^i \hat{s}\left(\hat{s}-m_Z^2\right)}
{\left(\hat{s}-m_Z^2\right)^2 + m_Z^2\Gamma_Z^2}.
 \end{equation}

The couplings $d_V$ and $d_A$ are defined as above and $h_{L/R}=
\frac{g_{L/R}g_Z}{e}$
where $g_{L/R}$ is the coupling to left/right-handed sleptons at the gauge 
boson-slepton-slepton vertex.

The $\sin^2\theta^*$ angular distribution contrasts with the asymmetric 
distribution for the heavy leptons as seen in eq.~(\ref{dxs}).  This will 
be studied in section~\ref{angdist}.

\section{Implementation}
\label{impl}

\subsection{Modifications to {\small HERWIG 6.3}}

The above formulae were incorporated into a new subroutine added to 
{\small HERWIG 6.3} together with new particle entries (for the heavy 
leptons) and new process codes (IPROCs) for the new processes.

Increasing the mass of the new leptons allowed the variation of cross section
with mass to be studied - the mass was varied over a range from of order the 
top quark mass to a maximum of order 1~TeV\@.  The lifetimes of the charged 
leptons produced were set to 1~s so that the number of decays occurring inside
the detector will be negligible.

The angular distribution of the produced leptons was studied over the same 
mass range to investigate the possibility of distinguishing heavy leptons 
from MSSM sleptons.

\subsection{Time-of-Flight Technique}
\label{tof}

Our main aim was to consider the possible detection of charged heavy leptons 
at the LHC and for what mass ranges this might be practical.  This work 
refers specifically to the technical specifications of the ATLAS detector 
\cite{ATLASTDR1,ATLASTDR2,ATLASmTDR}.  However we believe similar results will
be obtained for the CMS experiment \cite{CMS}.

The method used was a time-of-flight technique as discussed in 
ref.~\cite{H+P}.  This method utilises the fact that, when compared to 
relativistic particles, there is a considerable time delay for heavy particles
to reach the muon system.  Heavy charged leptons like those being considered 
in this work will be detected in both the central tracker and the muon 
chambers, and from the measured momentum and time delay it is possible to 
reconstruct the mass.

Imperfections in the time and momentum resolutions will 
cause a spread of the mass peak.  Uncertainty over which bunch crossing a 
particular detected particle comes from may also provide a background signal 
(from muons produced in Drell-Yan processes and also from heavy quark decays).
Plots from \cite{ATLASmTDR} display the behaviour of the processes which are 
most likely to produce muons which might be mis-identified as heavy leptons.  
These are reproduced in figures~\ref{muonpT} and \ref{muonbg}.

\DOUBLEFIGURE{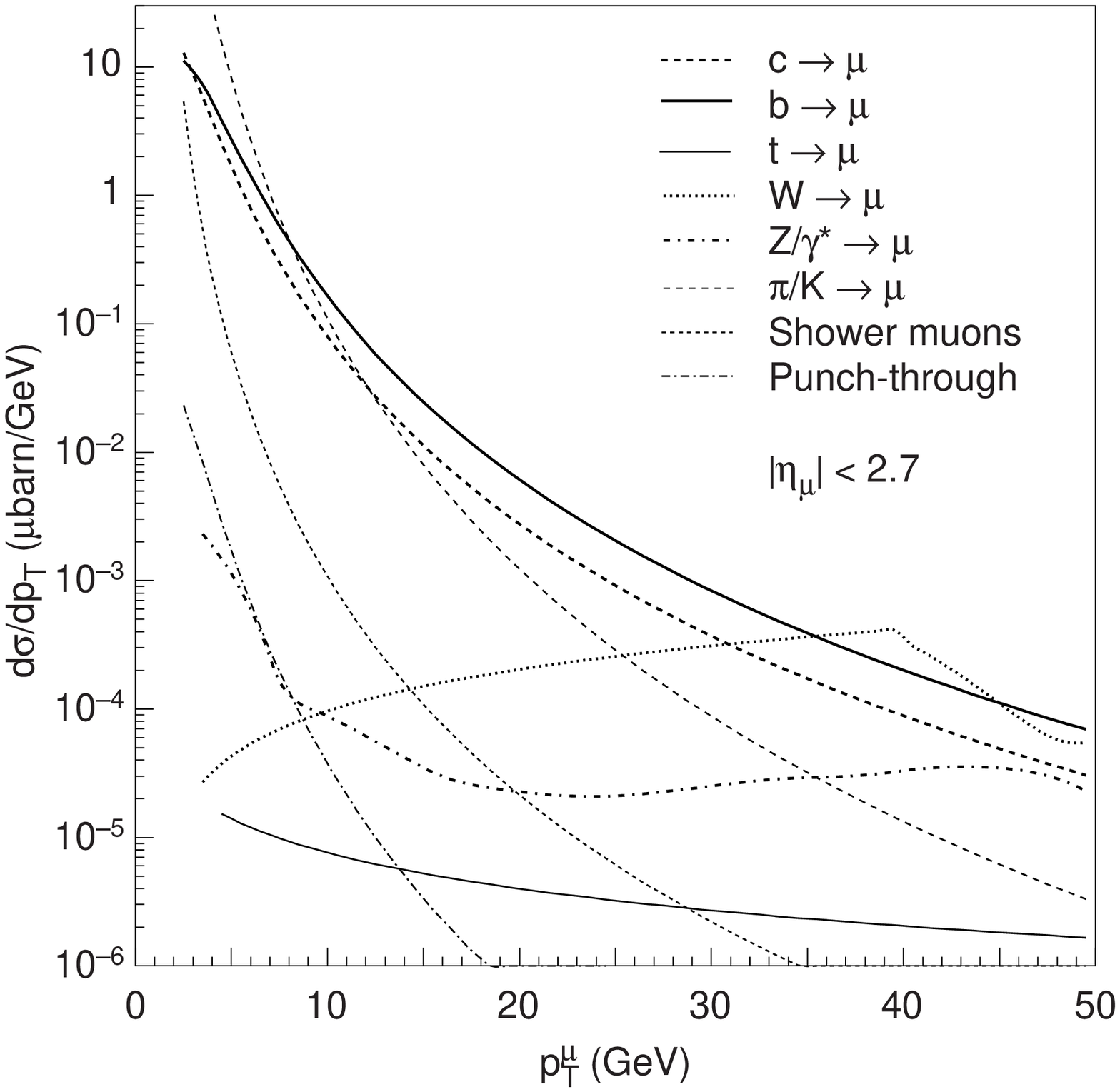,width=7.35cm}{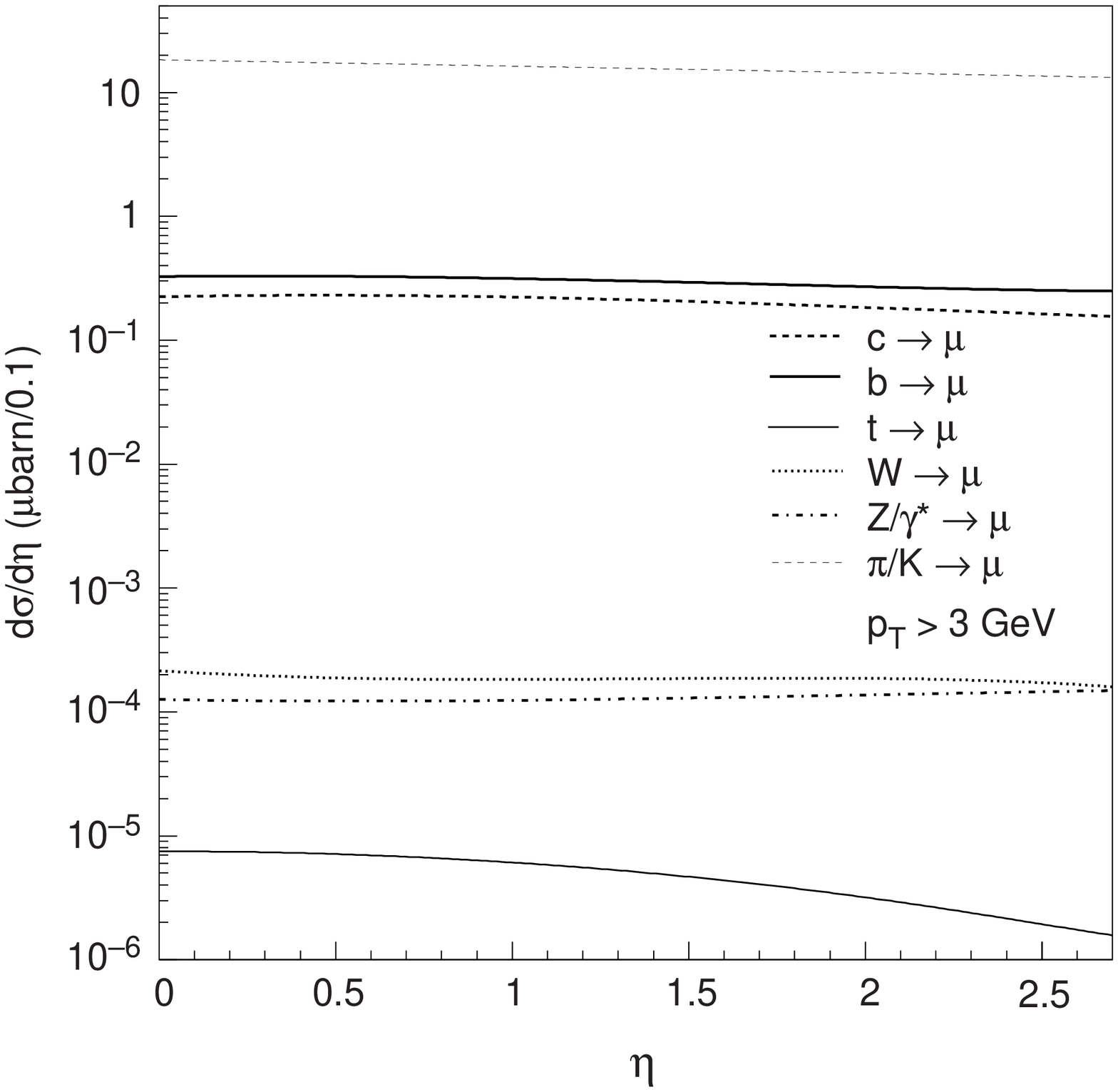,width=7.5cm}
{Transverse momentum dependence of the inclusive muon cross section 
integrated over $|\eta|< 2.7$.  The horizontal scale is the transverse 
momentum at production. (Reproduced from \cite{ATLASmTDR}).
\label{muonpT}}
{Rapidity-dependence of the inclusive muon cross sections, integrated
over \mbox{3 $<p_T<$ 50~GeV}. (Reproduced from \cite{ATLASmTDR}).
\label{muonbg}}

The studies presented here do not attempt to model the background, which it is
not thought will be significant (there is no ``physics'' background, only 
``detector'' background as discussed above), and use a simple cut (as in 
\cite{H+P}) on the range of time delays allowed.

The calculation of a reconstructed mass from the time delay $\Delta t$ and 
momentum is straightforward.  For a lepton hitting the radial part of the 
muon spectrometer, the time delay with respect to a relativistic 
($\beta=\frac{v}{c}=1$) particle is given by

\begin{equation}
\Delta t = \frac {r} {p_T} \left(E-p\right), 
\end{equation}

\noindent where $r$ is the radius of the outer layer of the muon system, 
$p_T$ the transverse momentum, and $E$ and $p$ the total energy and momentum 
respectively.  Substituting for $E$ in the energy-momentum invariant 
$E^2-p^2=m^2$ gives the result

\begin{equation}
m^2 = p^2 \left( \frac {1}{\beta^2} - 1 \right)
= \frac {p_T \Delta t} {r} \left(2p + \frac {p_T \Delta t} {r} \right).
\label{masssq}
\end{equation}

\noindent It is also necessary to check for occasions when the lepton hits the
endcap of the detector by considering the magnitude of $\frac {p_T} {p_z}$\@.
The muon system is modelled approximately as a cylinder of radius 10~m and a 
half-length of 20~m \cite{ATLASmTDR}.  A pseudo-rapidity cut requiring that 
$\eta<|2.7|$ was applied to take account of the region close to the beam where
particles cannot be detected.

From the expression (\ref{masssq}), the time delay and measured momentum for 
any particle detected in the muon system can be used to calculate the mass.  
As in \cite{H+P} the time delay was smeared with a Gaussian (although a width 
of 0.7~ns  \cite{ATLASmTDR} rather than 1~ns was used) and a cut on 
the smeared time delay was applied such that 
$10 \mbox{ ns} < \Delta t < 50\mbox{ ns}$.  Increasing the lower limit 
on $\Delta t$ would  
reduce the efficiency but improve the mass resolution by removing many of the 
high $\beta$ leptons for which the time resolution contribution is large.
The upper limit eliminates very slow particles which lose most of their energy
in the calorimeter.  An alternative upper limit of 25~ns, reflecting practical
concerns, is discussed in section \ref{masspeaks}.  

The most important contribution to the momentum uncertainty $\Delta p$ is due
to the measurement error on the sagitta (the deviation from a straight line of
a charged particle in a magnetic field).  $\Delta p_{sag}$ was taken into 
account by using 

\begin{equation}
\frac {\Delta p_{sag}}{p^2} = 1.1\times10^{-4}, 
\label{sagerror}
\end{equation}

\noindent where $p$ is the total momentum in GeV.  The constant in 
eq.~(\ref{sagerror}) was obtained from \cite{ATLASTDR1} as an average over the
different parts of the ATLAS detector.  This expression should be valid near 
the discovery limit for heavy leptons when multiple scattering is 
insignificant - the maximum time delay cut is found to remove most of the low 
$\beta$ leptons for which multiple scattering would have more effect.  However
to allow investigations over a full mass range a multiple scattering term 
($\Delta p_{ms}$) was also incorporated with  

\begin{equation}
\Delta p_{ms} = 2\times10^{-2}\sqrt{p^2 + m^2}.
\label{mserror}
\end{equation}

\noindent The constant depends on the material distribution in the 
spectrometer - the above value is given in \cite{ATLMUON} for the ATLAS muon 
detector.

The two errors in eqs.~(\ref{sagerror}) and (\ref{mserror}) were taken to be 
independent and hence combined in quadrature.  A third possible contribution 
to momentum resolution - due to fluctuation in the energy loss in the 
calorimeter - is neglected as it is always dominated by one of the other 
terms.

Using the momentum and time resolution as described above we are able to 
satisfactorily reproduce the mass resolution as a function of $\beta$ for 
101~GeV particles as given in \cite{ATLASTDR1}.

The efficiency of the muon detection system is approximated to be 85\%,
independent of the particle momentum \cite{ATLASTDR1}.

\section{Results}
\label{results}

\subsection{Backgrounds}
\label{bg}

Any ``detector'' background is greatly dependent on the 
details of the muon detection system.  Such a background would come from muons 
either from the decay of heavy quarks or from Drell-Yan (both these processes 
have very high cross sections compared to that for heavy lepton production).  
If timing inadequacies do mean that some of these muons are mistakenly thought
to have come from an earlier interaction \emph{i.e.} they appear to have a 
large time delay, they could obscure the exotic lepton mass peak. For a 
background signal to look like a heavy lepton neutral current signal, however,
two opposite charge muons would have to be mis-identified at the same time 
which makes it very unlikely that background could be significant. 

To eliminate any possible background from heavy quark decays, a cut requiring 
the transverse momentum to be greater than 50~GeV was made to produce the 
plots in section~\ref{masspeaks}.  Figures~\ref{bp} and \ref{p250} show that a
$p_T$ cut at 50~GeV should remove most of the muons from bottom 
quark decay without significantly reducing the exotic heavy lepton signal.  It
can be seen that as the mass of the exotic heavy lepton increases, the peak in 
the $p_T$ spectrum becomes less well defined and shifts to higher values.  Top
quark decays can give rise to muons at high $p_T$ but the small top 
cross~section, combined with the improbability of mis-identifying two muons,
should suffice to eliminate this source of background. 

\FIGURE{
\scalebox{0.5}{\rotatebox{-90}{\includegraphics[width=\textwidth]{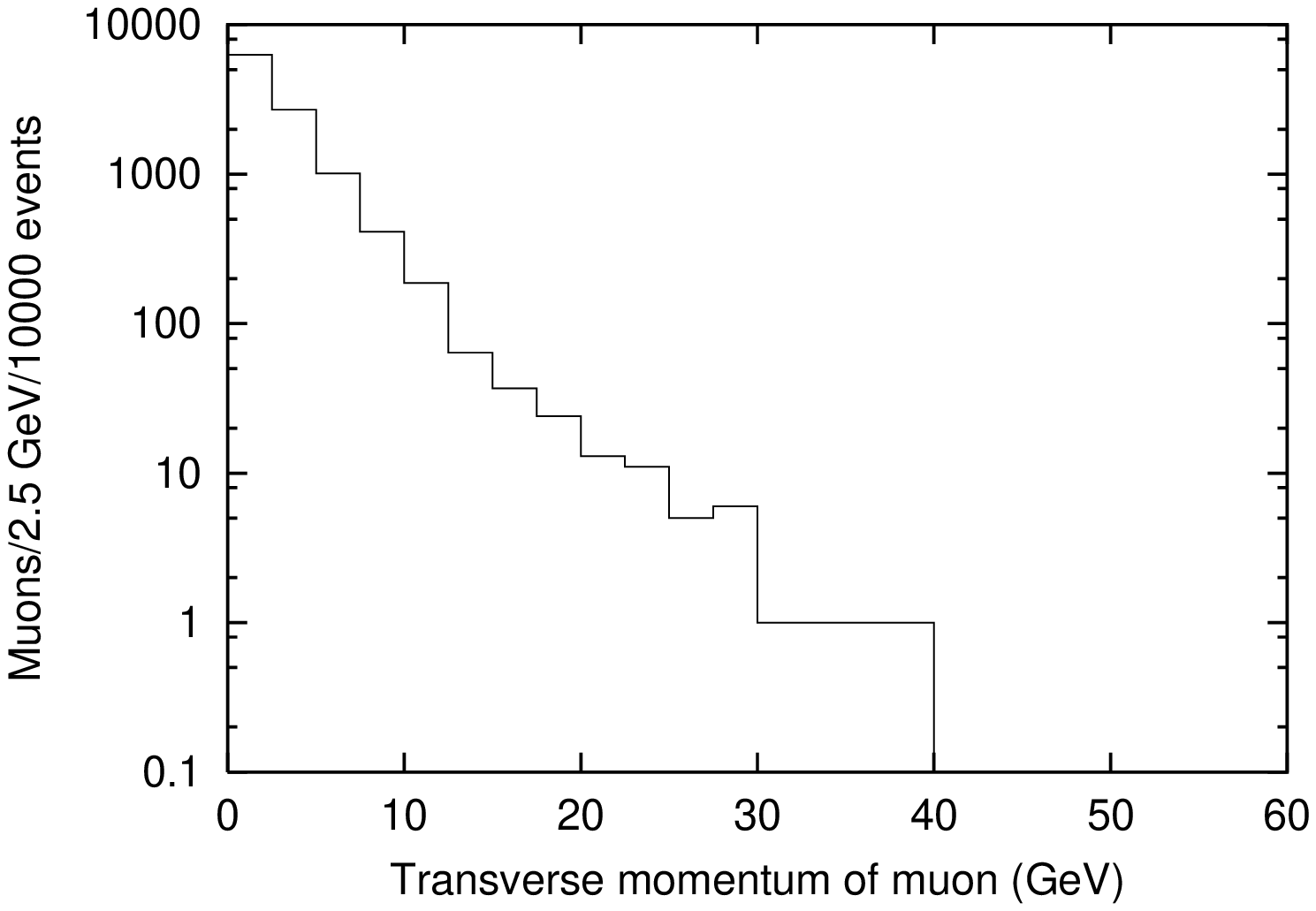}}}
\caption{$p_T$ spectrum for muons produced in b decays.}
\label{bp}
}

\FIGURE{
\unitlength1cm
\begin{minipage}[t]{7.0cm}
\scalebox{0.57}{\rotatebox{-90}{\includegraphics
{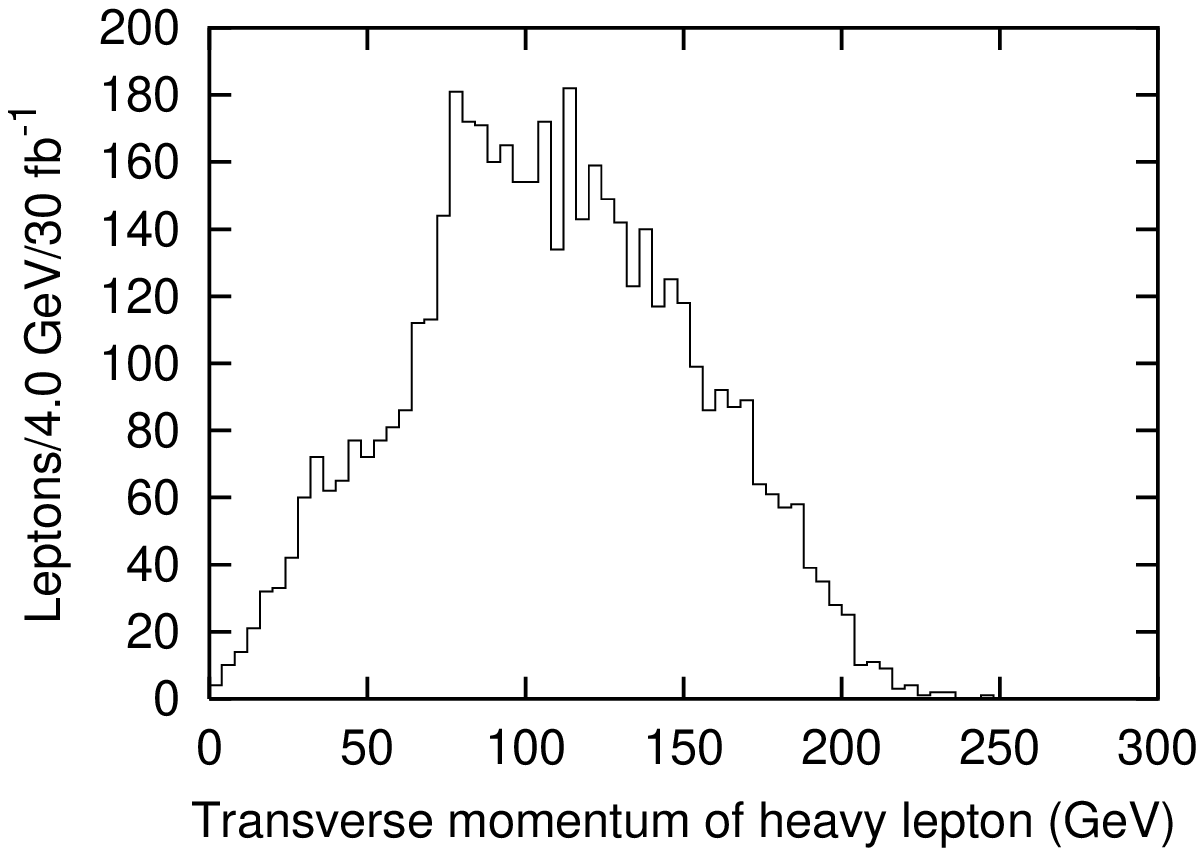}}}
\end{minipage}
\hfill
\begin{minipage}[t]{7.0cm}
\scalebox{0.57}{\rotatebox{-90}{\includegraphics
{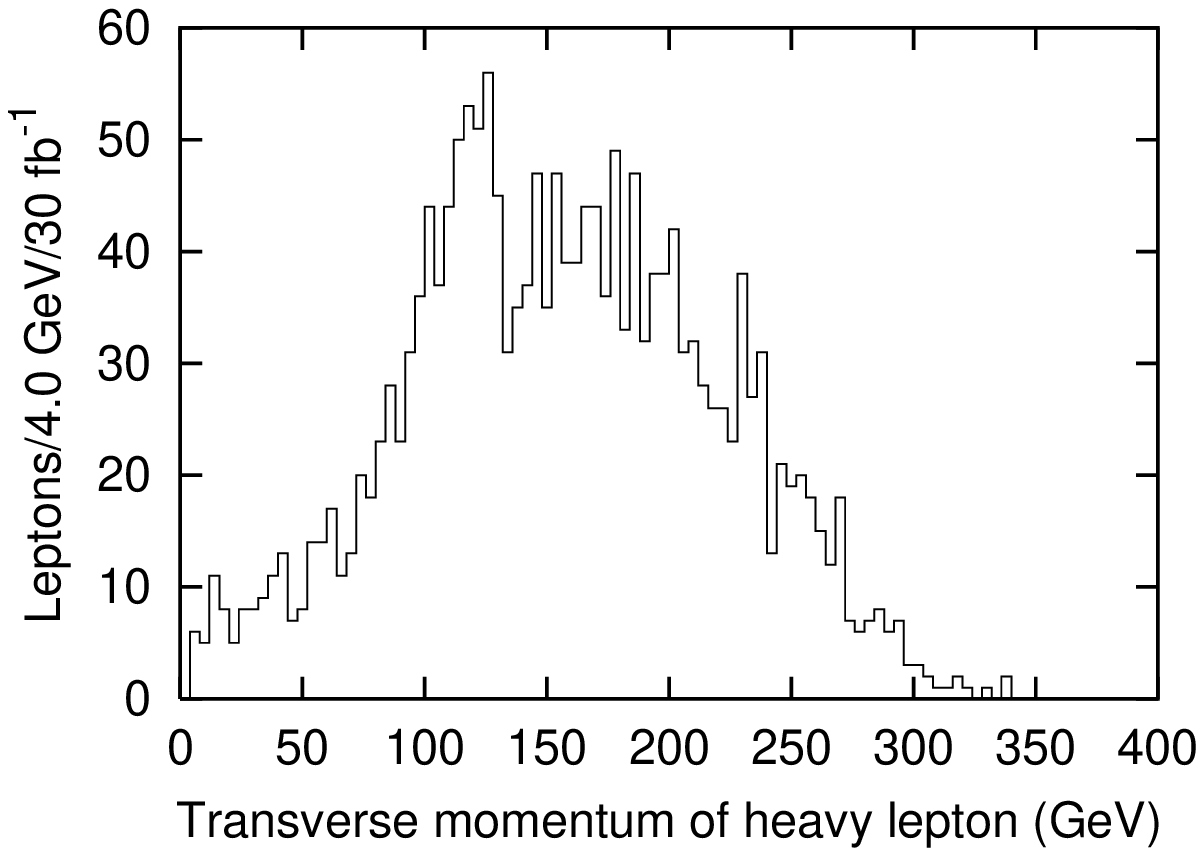}}}
\caption{$p_T$ spectra for 175 and 250~GeV heavy leptons from Drell-Yan 
processes.
\label{p250}}
\end{minipage}
}

\subsection{Heavy Lepton Mass Peaks}
\label{masspeaks}

The results presented in this section show the reconstructed mass peaks for 
an integrated luminosity of 100~fb$^{-1}$ after the cuts outlined above have 
been applied. 

The possibility of detection at the LHC is found to be entirely cross~section 
limited - the decrease of cross~section with increasing lepton mass is shown
in figure~\ref{cross}.  The fraction of leptons passing the cuts actually 
slightly increases with mass due to the longer time delays.

\FIGURE{
\scalebox{0.5}{\rotatebox{-90}{\includegraphics[width=\textwidth]
{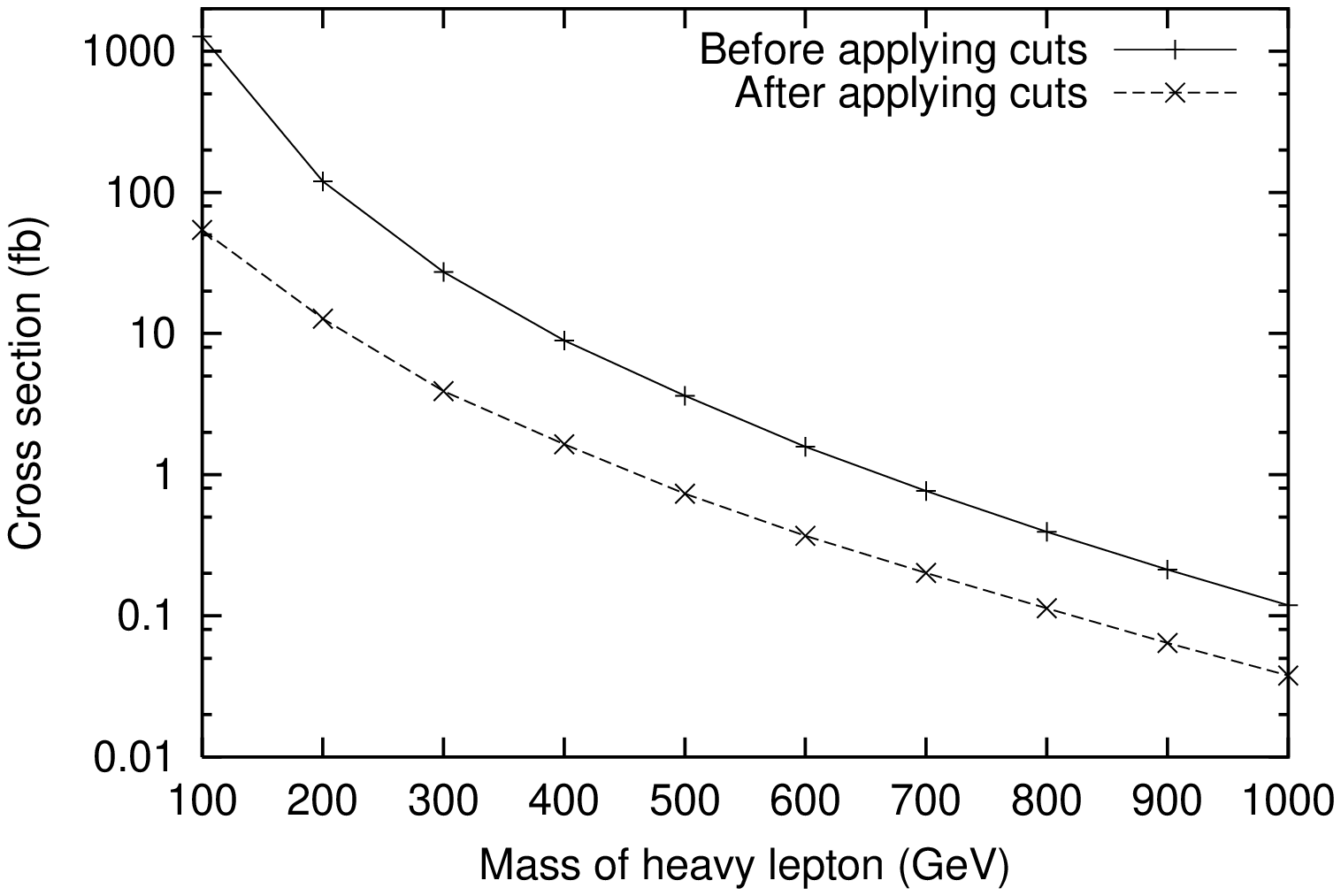}}}
\caption{Cross~section for Drell-Yan heavy lepton production as a function of
the lepton mass.}
\label{cross}
}

Figure~\ref{1000} for the neutral Drell-Yan process makes it
clear that for masses up to about 1~TeV it should be possible to detect new
charged heavy leptons at the LHC (particularly if a larger integrated 
luminosity can be obtained).  A discovery criterion of 10 leptons (from 5 
events) in the mass peak is found to give a mass limit of 950~GeV for the 
detection of such leptons using an integrated luminosity of 100~fb$^{-1}$.

To reduce possible backgrounds the plots are based on events from 
which both leptons produced passed the cuts imposed.

\FIGURE{
\unitlength1cm
\begin{minipage}[t]{7.0cm}
\scalebox{0.57}{\rotatebox{-90}{\includegraphics
{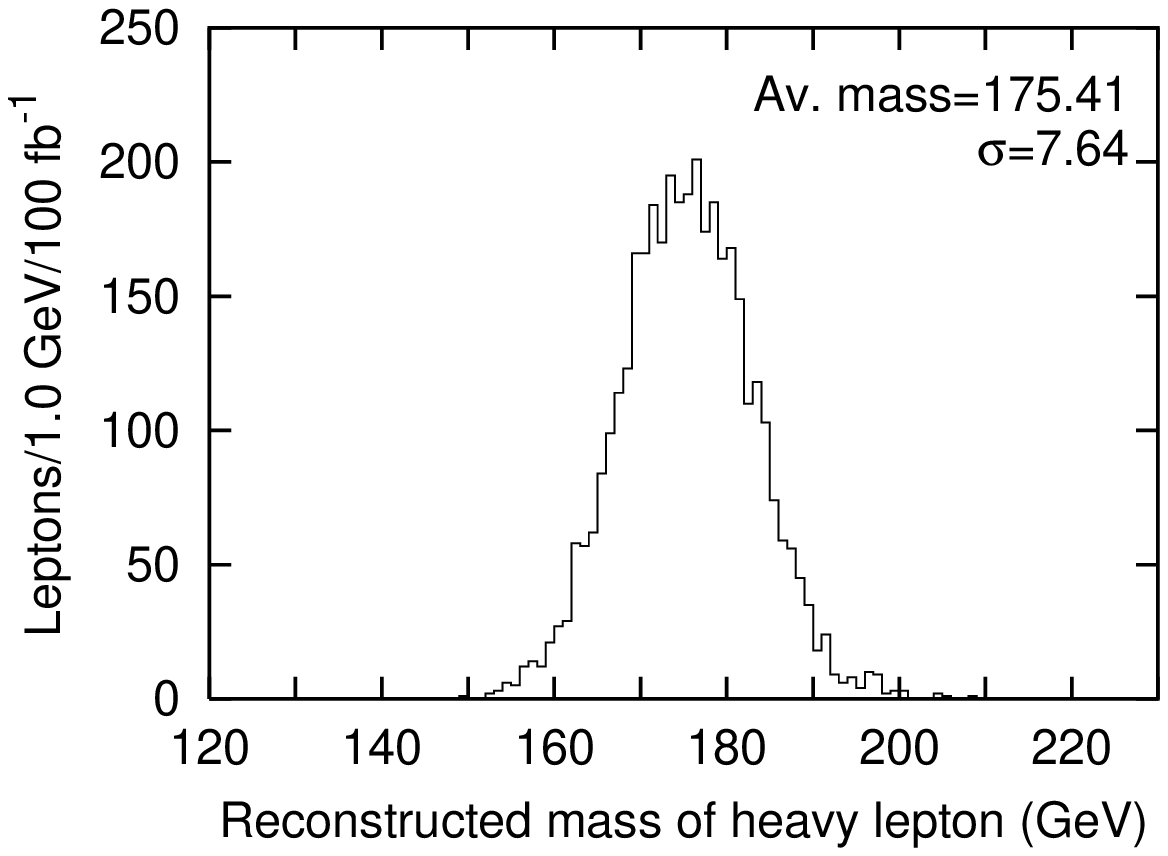}}}
\end{minipage}
\hfill
\begin{minipage}[t]{7.0cm}
\scalebox{0.57}{\rotatebox{-90}{\includegraphics
{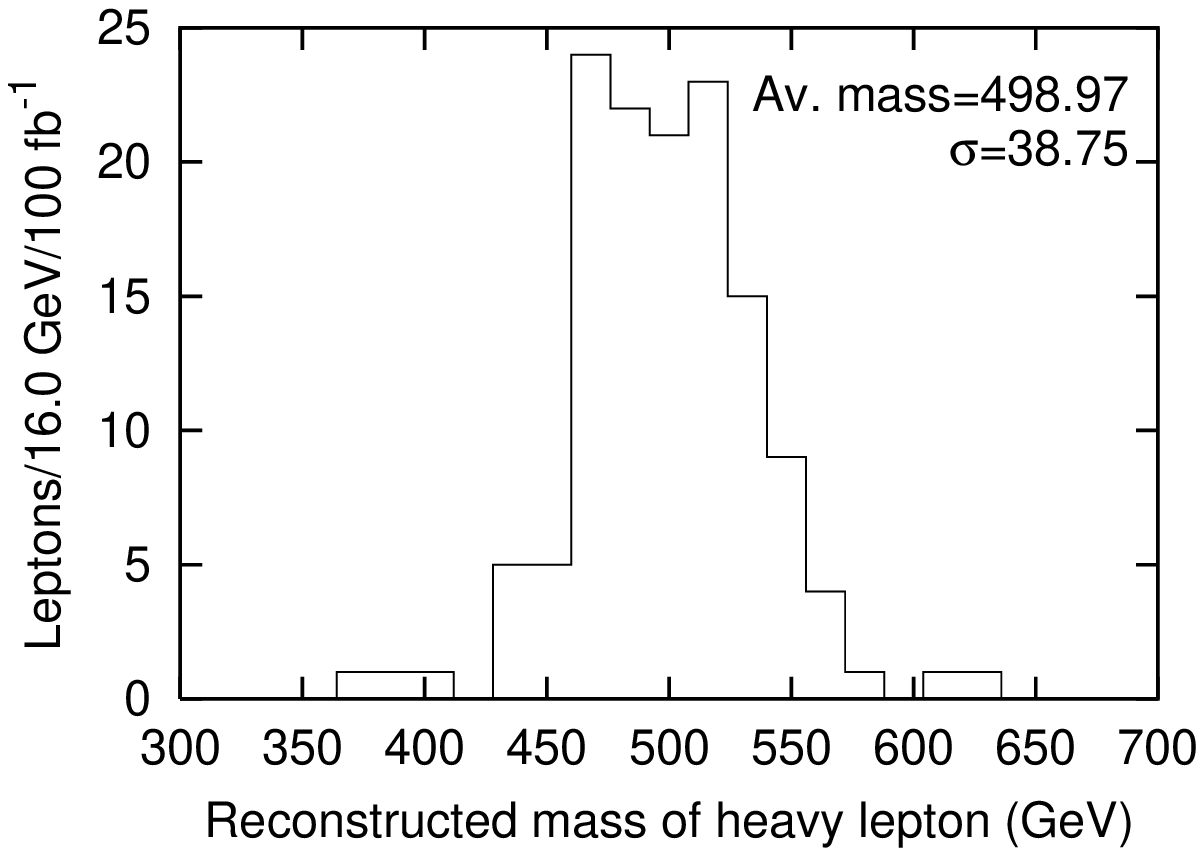}}}
\end{minipage}

\unitlength1cm
\begin{minipage}[t]{7.0cm}
\scalebox{0.57}{\rotatebox{-90}{\includegraphics
{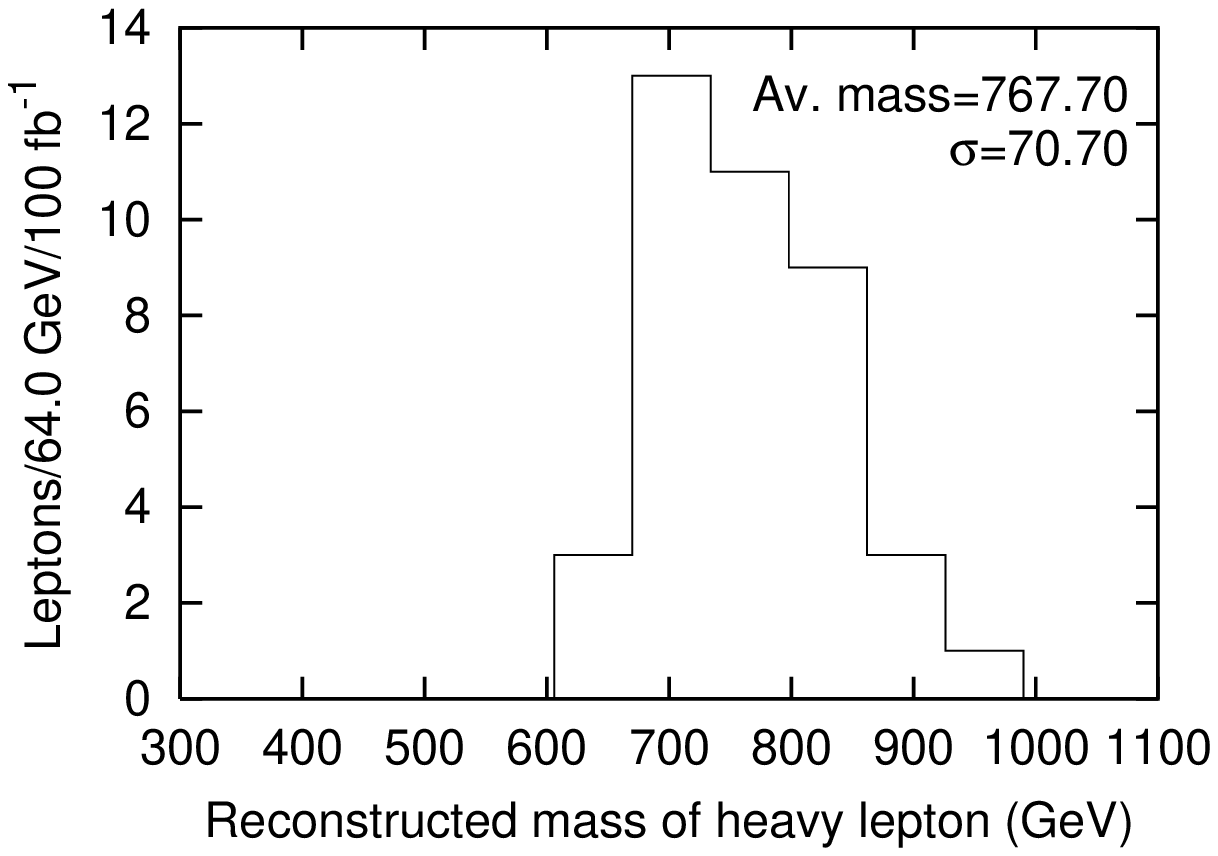}}}
\end{minipage}
\hfill
\begin{minipage}[t]{7.0cm}
\scalebox{0.57}{\rotatebox{-90}{\includegraphics
{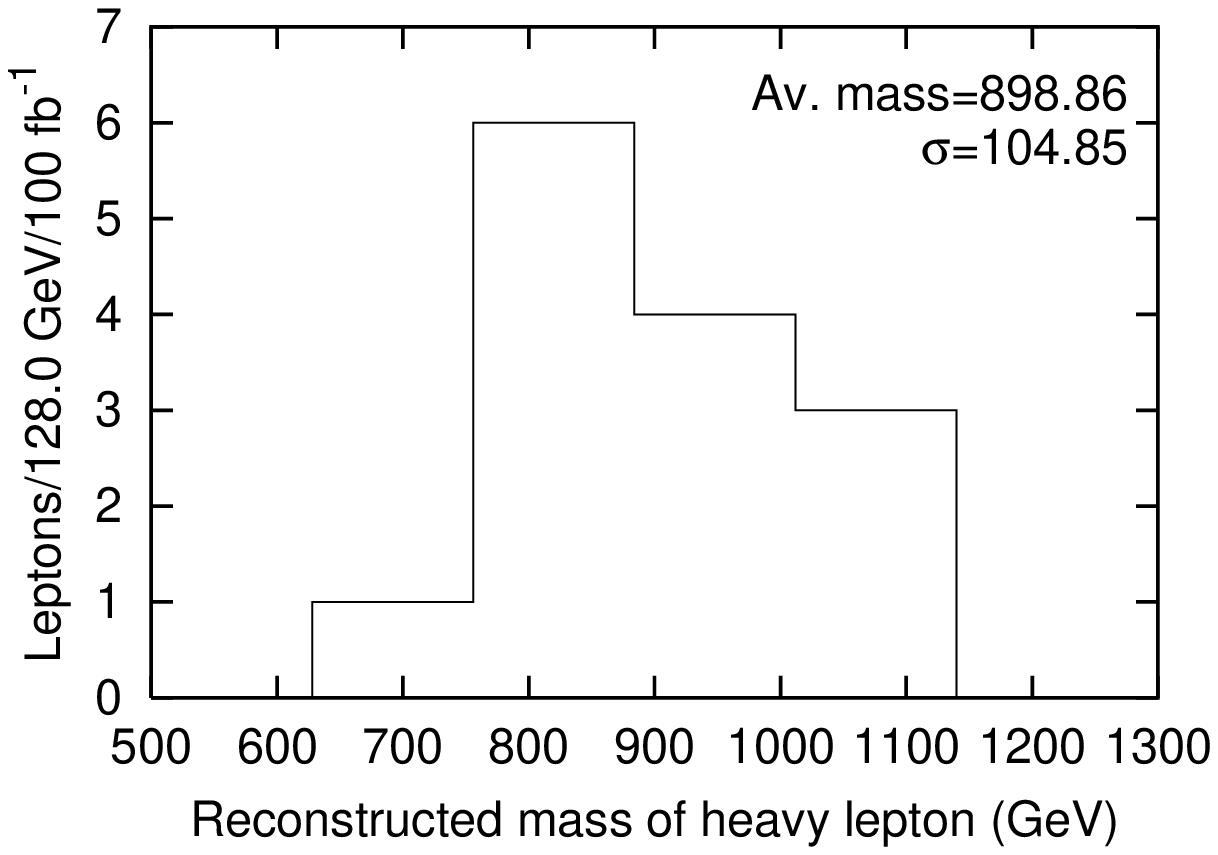}}}
\caption{Reconstructed mass peaks for exotic leptons with masses of
175~GeV, 500~GeV, 750~GeV and 900~GeV\@.
\label{1000}}
\end{minipage}
}

We note that the long drift times in the muon chambers used by ATLAS
allow the recording of events with time delays at arrival much greater
than the 25 ns beam crossing interval of the LHC \cite{ATLASmTDR}. We have 
assumed a maximum time delay of 50 ns following \cite{H+P}. However, to 
correctly associate the delayed track in the muon system with the event 
recorded in the other detector systems would require a specialised trigger,
based, for example, on the presence of high $p_T$ tracks in the inner
detector, which would identify the true event time. Without this, the
maximum delay time allowable would be 25 ns, and the discovery reach
would be reduced to a lepton mass of 800~GeV. The statistical
sample available for the angular distribution analysis in section 
\ref{angdist} would also be reduced by a factor of about 2, depending on the 
lepton mass.

\subsection{Distinguishing Leptons from Sleptons}
\label{angdist}

As mentioned in section~\ref{HERWIG} it should be possible to distinguish 
these new heavy leptons from heavy scalar leptons by studying the angular 
distribution with which they are produced.  The two different angular 
distributions in the centre-of-mass frame are shown in eqs.~(\ref{dxs}) and
(\ref{sdxs}).  The application of the cuts described in section~\ref{tof} as 
well as the effects of the mass and momentum resolution mean that the 
observed angular distributions are somewhat different to these.

The forward-backward asymmetry for the exotic heavy leptons was not considered
because of the difficulty in a $pp$ collider of distinguishing the quark and 
anti-quark in the centre-of-mass frame.  Hence only $|\cos\theta^*|$ was used
when the angular distributions were compared.  Figure \ref{a400} shows the 
angular distribution for both exotic leptons and sleptons of mass 400~GeV\@.

\FIGURE{
\scalebox{0.5}{\rotatebox{-90}{\includegraphics[width=\textwidth]
{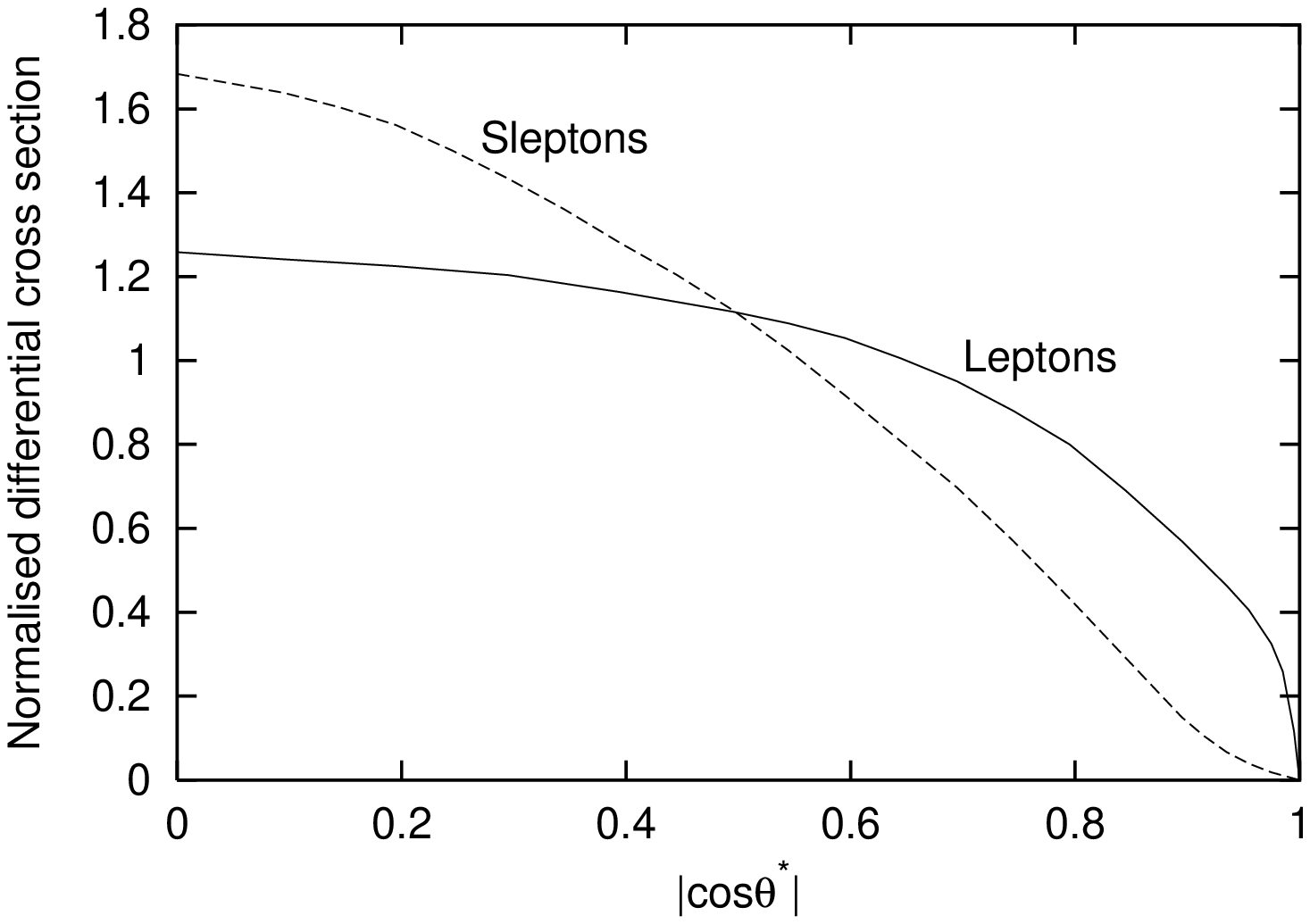}}}
\caption{Angular distributions of detected pairs of 400 GeV particles.
$\theta^*$ is the angle between the outgoing particle and incoming quark 
directions}
\label{a400}
}

A chi-squared test was applied to find the probability that a scalar model 
describes the data.  The model was the angular distribution of a large Monte 
Carlo sample of scalars, and the total number of detected scalar pairs was 
rescaled to the expected number of detected pairs of heavy leptons.  An 
``average'' data-set was obtained from a rescaled Monte Carlo sample of 
leptons, generated according to their angular distribution.  Both these 
rescaled distributions had negligible theoretical errors compared to the 
(Poisson distributed) statistical errors on the model for 100~fb$^{-1}$ of
integrated luminosity.

Figure~\ref{prob} shows that it will be possible to rule out scalar leptons 
at a confidence level of better than 90\% up to a mass of 580~GeV\@.  The 
angular distribution does not change significantly over the mass range of this
plot - the decreasing probability of ruling out the scalar hypothesis as the
mass increases is mainly because of the decreasing statistics.

\FIGURE{
\scalebox{0.5}{\rotatebox{-90}{\includegraphics[width=\textwidth]{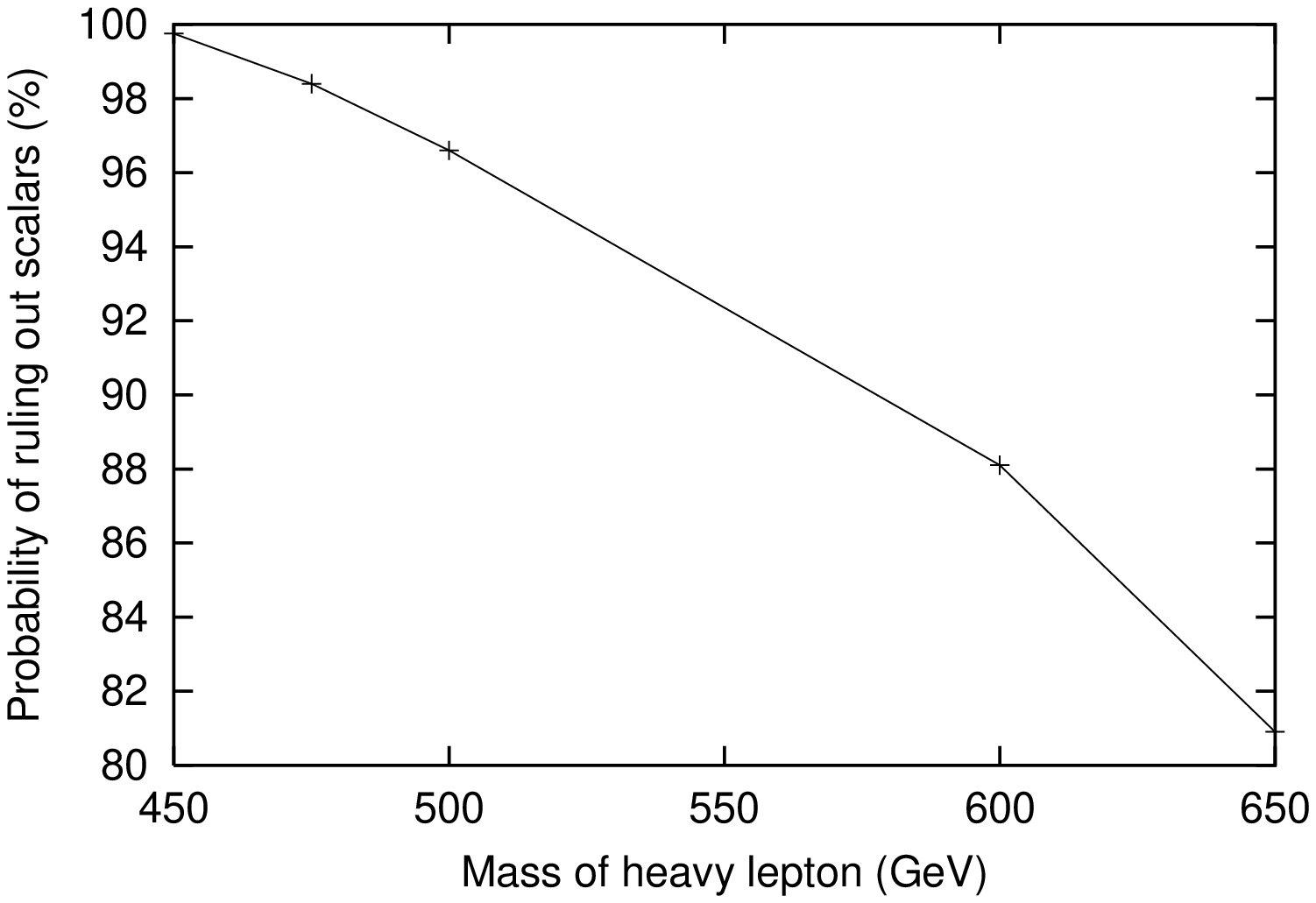}}}
\caption{Probability at which it will be possible to rule out the scalar
hypothesis by studying the angular distribution of the detected leptons.}
\label{prob}
}

\section{Conclusions}

This work was motivated by some of the new models of physics beyond the 
Standard Model in which the gauge couplings unify at an intermediate energy
scale ($\sim 10^{11}$~GeV)\@.  Many of these intermediate scale models include
additional leptons added to the Minimal Supersymmetric Standard Model and 
which are expected to have TeV scale masses.

We conclude that, assuming Standard Model couplings and a long enough 
lifetime, it should be possible to detect the charged heavy leptons in 
intermediate scale models up to masses of 950~GeV with 100~fb$^{-1}$ of 
integrated luminosity.  Background from mis-identified muons should be 
negligible and can be reduced by the application of a 
$p_T^{\mbox{\tiny{min}}}$ cut at about 50~GeV\@.  It will be difficult to use 
the angular distribution of the produced particles to distinguish them from 
scalar leptons for masses above 580~GeV because the falling cross section 
limits the available statistics.   

The absence of a specialised trigger in the inner detector could reduce the 
discovery limit to 800~GeV and would reduce the statistics available for the 
angular distribution analysis.

No present cosmological or experimental limits were found to rule out 
additional exotic leptons at the masses considered in this work. 

\acknowledgments

We thank members of the Cambridge SUSY Working Group, particularly 
F.~Quevedo and D.~Grellscheid, for helpful discussions.  This work was funded 
by the U.K. Particle Physics and Astronomy Research Council.

\end{document}